%% file: main.tex
\documentclass[conference]{IEEEtran}
\IEEEoverridecommandlockouts
\usepackage{cite}
\usepackage{amsmath,amssymb,amsfonts}
\usepackage{algorithmic}
\usepackage{graphicx}
\usepackage{textcomp}
\usepackage{xcolor}
\usepackage{enumitem}
\usepackage{array}
\usepackage{soul}
\usepackage{subcaption}
\usepackage{lipsum}
\usepackage[font=small]{caption}
\newcommand{\PreserveBackslash}[1]{\let\temp=\\#1\let\\=\temp}
\newcolumntype{C}[1]{>{\PreserveBackslash\centering}p{#1}}
\newcolumntype{R}[1]{>{\PreserveBackslash\raggedleft}p{#1}}
\newcolumntype{L}[1]{>{\PreserveBackslash\raggedright}p{#1}}
\usepackage[hidelinks]{hyperref}

\def\BibTeX{{\rm B\kern-.05em{\sc i\kern-.025em b}\kern-.08em
    T\kern-.1667em\lower.7ex\hbox{E}\kern-.125emX}}
\begin{document}

\title{
RAD-Sim: Rapid Architecture Exploration for Novel Reconfigurable Acceleration Devices
}

\author{
\IEEEauthorblockN{Andrew Boutros$^{1,2}$, Eriko Nurvitadhi$^{2}$ and Vaughn Betz$^{1}$}
\IEEEauthorblockA{$^{1}$University of Toronto and Vector Institute for AI ~~~~ $^{2}$Programmable Solutions Group, Intel Corporation\\
E-mails: andrew.boutros@mail.utoronto.ca, eriko.nurvitadhi@intel.com, vaughn@eecg.utoronto.ca}
}


\maketitle

\begin{abstract}
With the continued growth in field-programmable gate array (FPGA) capacity and their incorporation into new environments such as datacenters, we have witnessed the introduction of a new class of reconfigurable acceleration devices (RADs) that go beyond conventional FPGA architectures. 
These devices combine a reconfigurable fabric with coarse-grained domain-specialized accelerator blocks all connected via a high-performance packet-switched network-on-chip (NoC) for efficient system-wide communication. 
However, we lack the tools necessary to efficiently explore the huge design space for RADs, study the complex interactions between their different components and evaluate various combinations of design choices. 
In this work, we develop RAD-Sim, a cycle-level architecture simulator that allows rapid application-driven exploration of the design space of novel RADs. 
To showcase the capabilities of RAD-Sim, we map and simulate a state-of-the-art deep learning (DL) inference overlay on a RAD instance incorporating an FPGA fabric and a complex of hard matrix-vector multiplication engines, communicating over a system-wide NoC.
Through this example, we show how RAD-Sim can help architects quantify the effect of changing specific architecture parameters on end-to-end application performance.
\end{abstract}

\begin{IEEEkeywords}
FPGA, NoC, accelerator blocks, architecture simulator, deep learning
\end{IEEEkeywords}

\fontsize{10pt}{11.5pt}\selectfont

\input{01_introduction}
\input{02_background}
\input{03_radsim}
\input{04_npu}

\section{Conclusion}

As FPGAs continue to grow in capacity and move into datacenters, there is demand for both faster time-to-solution and increased acceleration of key workloads. 
These pressures are producing a shift towards novel RADs that combine the hardware reconfigurability of FPGAs with domain specific accelerator blocks and NoCs for full-featured system-wide communication. 
However, the tools required for the exploration of the huge design space of such devices do not exist.
In this work, we introduce RAD-Sim, a SystemC-based application-driven simulator that can be used for rapid architecture exploration of RADs incorporating conventional FPGAs, high-performance packet-switched NoCs, and coarse-grained hard accelerator blocks.
This cycle-level simulator enables studying different RAD architectures and quantifying the effect of specific design choices on end-to-end application performance.
To showcase the capabilities of RAD-Sim, we present an example design that maps the state-of-the-art NPU DL inference overlay on an example RAD instance.
Both RAD-Sim and the NPU example design are open source so that the research community can leverage them to drive further innovations in RAD architecture.

\section*{Acknowledgements}
The authors would like to thank the Intel/VMware Crossroads 3D-FPGA Academic Research Center and the NSERC/Intel Industrial Research Chair in Programmable Silicon for funding support.

\bibliographystyle{./bibliography/IEEEtran}
\bibliography{ref.bib}

\end{document}

%% file: 01_introduction.tex
\section{Introduction}

\begin{figure}[t!]
    \centering
    \includegraphics[width=0.8\linewidth]{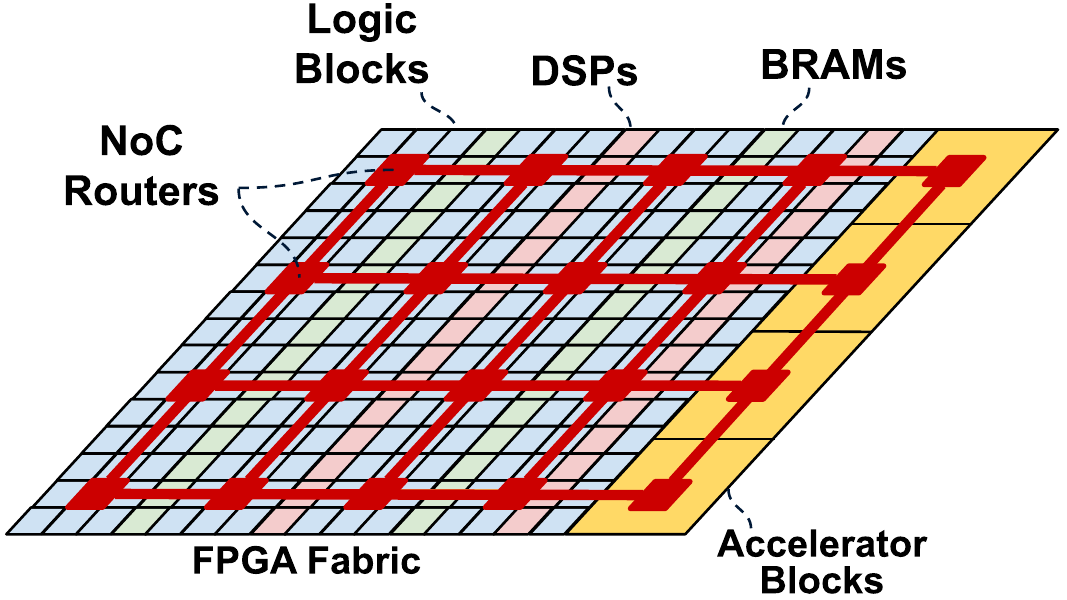}
    \caption{Example RAD instance incorporating a conventional FPGA fabric, a side complex of coarse-grained accelerator blocks, and a packet-switched hard NoC for system-wide communication.}
    \label{fig:rad-instance}
    \vspace{-0.2cm}
\end{figure}

Field-programmable gate arrays (FPGAs) have evolved significantly over the past thirty years from simple arrays of reconfigurable logic and routing into complex heterogeneous devices with on-chip memories (BRAMs), digital signal processing blocks (DSPs), and high-speed transceivers~\cite{boutros2021fpga}.
More recently, we have witnessed the emergence of \textit{beyond-FPGA} reconfigurable acceleration devices (RADs). 
These devices combine a conventional FPGA fabric with a number of coarse-grained application-specific accelerator blocks, communicating via high-performance networks-on-chip (NoCs) as depicted in Fig.~\ref{fig:rad-instance}; an exemplar is the Xilinx Versal architecture~\cite{gaide2019xilinx}.
With advances in multi-die integration, RADs can also span multiple dice with the system-level NoC(s) acting as a continuous communication plane between them. 

The combination of these different components in a RAD results in a huge design space, opening up a myriad of research questions on how we should architect these devices given the complex interactions between their different components.
Although FPGA fabric architecture has been extensively studied for many years, the tools and methodologies for exploring and evaluating fabric architectures are inadequate for architecture exploration of novel RADs. 
Firstly, they evaluate candidate fabric architectures based on application-agnostic performance metrics such as the maximum operating frequency of benchmark circuits.
For RADs with coarse-grained accelerator blocks and latency-insensitive NoC communication, performance metrics used must go beyond the operating frequency of the logic implemented on the FPGA fabric and capture end-to-end application performance. 

Secondly, FPGA architecture exploration flows are mainly driven by benchmarks written in hardware description language (HDL) and rely on register-transfer level (RTL) simulation for functional verification.
This requires developing a tremendous amount of RTL infrastructure for both applications and system components such as the NoC routers and hard accelerator blocks to perform system-level simulations for functional verification and performance estimation.
Such a slow and labor-intensive flow precludes broad exploration of RAD architectures and also limits the ability of architects to co-optimize applications and RAD platforms.
Finally, RAD architecture exploration tools need to evaluate new metrics such as the NoC traffic and congestion for different applications on a proposed architecture.

In this work, we first introduce RAD-Sim, a system-level application-driven architecture simulator for novel RADs that incorporate different NoCs, accelerator blocks, and fabric modules.
RAD-Sim takes as inputs a high-level SystemC description of application modules and accelerator blocks along with RAD architecture parameters, NoC specifications and router placement constraints.
It performs system-level simulation and produces end-to-end application performance and NoC traffic reports. 
It can also be used for functional verification of applications implemented on a given RAD instance when provided with user-specified test inputs and expected outputs.
We then present an example design to showcase the capabilities of RAD-Sim by mapping a state-of-the-art deep learning (DL) inference FPGA overlay, the neural processing unit (NPU), to an example RAD instance incorporating an FPGA, hard matrix-vector multiplication accelerator blocks, and a system-level NoC. 
Our contributions in this work are:

\noindent
\begin{itemize}[noitemsep,topsep=0pt,leftmargin=2\labelsep]
\item RAD-Sim, an open-source tool\footnote{Code can be downloaded at: \url{https://github.com/andrewboutros/rad-flow}} for rapid architecture exploration of novel RADs incorporating FPGA fabrics, accelerator blocks, and system-level NoCs.
\item An example design from the DL domain showing how RAD-Sim can help architects quantify the effect of different design choices on end-to-end application performance.
\end{itemize}

%% file: 02_background.tex
\section{Background and Related Work}
\label{sec:background}

\subsection{The Emergence of RADs}
In many FPGA datacenter deployments, the FPGA lies at the crossroads of data moving between different server endpoints.
The Microsoft Catapult v2 project \cite{caulfield2016cloud} places an FPGA as a \textit{bump-in-the-wire} between the network and server CPUs.
In this scenario, different network functionalities (e.g.~packet processing and cryptography) can be offloaded to the FPGA to free up CPU resources.
In addition, the network-connected FPGAs form a homogeneous datacenter-scale acceleration plane that can be flexibly reconfigured to accelerate different key datacenter applications such as DL workloads~\cite{fowers2018configurable}. 
In these deployments, the FPGA value comes not just from its reconfigurable logic, but also from its high-bandwidth I/Os.

However, the continuously increasing data flow of key workloads stresses the fine-grained programmable routing fabric especially when the FPGA is connected to several high-bandwidth external interfaces.
Prior work has shown that hardening packet-switched NoCs can mitigate these on-chip bandwidth challenges \cite{yazdanshenas2018interconnect,abdelfattah2016design}.  
Additionally, some compute operations in key applications are common across many workloads and their efficiency can be increased significantly by hardening them as coarse-grained accelerator blocks.
Taking DL acceleration as an example, the composition of layers, data manipulation between them, vector operations, and pre/post-processing stages might significantly differ between different workloads.
However, all of them include a large number of dot-product operations that can be hardened in the form of high-performance tensor cores for increased efficiency~\cite{langhammer2021stratix}.

As a result of these trends, we have started to witness the emergence of beyond-FPGA RADs that combine the flexibility of FPGAs, the efficiency of hard NoCs for data steering, and the high-performance of specialized accelerator blocks.
The Xilinx Versal architecture is an example of a RAD combining a conventional reconfigurable FPGA fabric, general-purpose ARM cores, and vector processors for DL acceleration, all communicating via a system-wide NoC \cite{gaide2019xilinx}.

\subsection{Conventional FPGA Architecture Exploration Flow}

Tools for FPGA architecture exploration, such as VTR \cite{murray2020vtr}, are well-established in the FPGA research community.
A typical FPGA architecture exploration flow consists of three main components:
(1) a suite of benchmark circuits that represent key FPGA application domains \cite{murray2013titan,arora2021koios};
(2) an architecture description defining the FPGA blocks, routing architecture, and their area/delay models; and
(3) a re-targetable CAD system that can map the given set of benchmarks to the specified FPGA architecture and produce area, timing, and power metrics.
This flow focuses only on the design of FPGA fabrics, primarily informed by application-agnostic metrics such as the maximum operating frequency of a benchmark circuit or the area cost of low-level FPGA circuitry.
This is not sufficient to explore and evaluate RAD architectures that include other complex components (e.g. NoCs and hard accelerator blocks), nor can it produce key system-level information such as NoC congestion and application throughput. 
NoC simulators also exist \cite{jiang2013detailed}, but as they lack features to simulate a coupled FPGA fabric, they also cannot fully evaluate a RAD.

\subsection{Architecture Simulators}

Architecture simulators are widely used to perform fast architecture exploration for classic von Neumann architectures as well as emerging compute technologies.
For example, the \texttt{gem5} \cite{binkert2011gem5} simulator performs high-fidelity cycle-level modeling of modern CPUs and can run full applications for different instruction set architectures.
GPGPU-Sim \cite{khairy2020accel} is another academic simulator for contemporary Nvidia GPU architectures that can run CUDA or OpenCL workloads and supports advanced features such as TensorCores and CUDA dynamic parallelism.
SIAM \cite{krishnan2021siam} is a more recent simulator focusing on emerging chiplet-based in-memory compute for deep neural networks. 
It integrates architecture, NoC, network-on-package, and DRAM models to simulate an end-to-end system.
In addition, specialized architecture simulators are commonly built to evaluate custom accelerator architectures such as in \cite{albericio2016cnvlutin,angizi2019mrima,yan2020hygcn}.
Our work, RAD-Sim, shares the same application-driven architecture exploration methodology of all these simulators but focuses on the reconfigurable computing domain.
Unlike other simulators like \texttt{gem5} or GPGPU-Sim, to evaluate RAD architectures, the input to the simulator is not just compiled application instructions.
Instead it can be a mix of instructions for any software-programmable coarse-grained accelerator blocks and custom user-defined modules implemented on the FPGA fabric. Another key difference is that both the placement of compute modules and their attachment to NoC routers are flexible (i.e. programmed at application design time) due to the FPGA reconfigurability.

%% file: 03_radsim.tex
\section{RAD Architecture Exploration Flow}
\label{sec:rad-sim}

\subsection{Flow Overview}
\label{sec:flow-overview}

\begin{figure}[t!]
    \centering
    \includegraphics[width=0.9\linewidth]{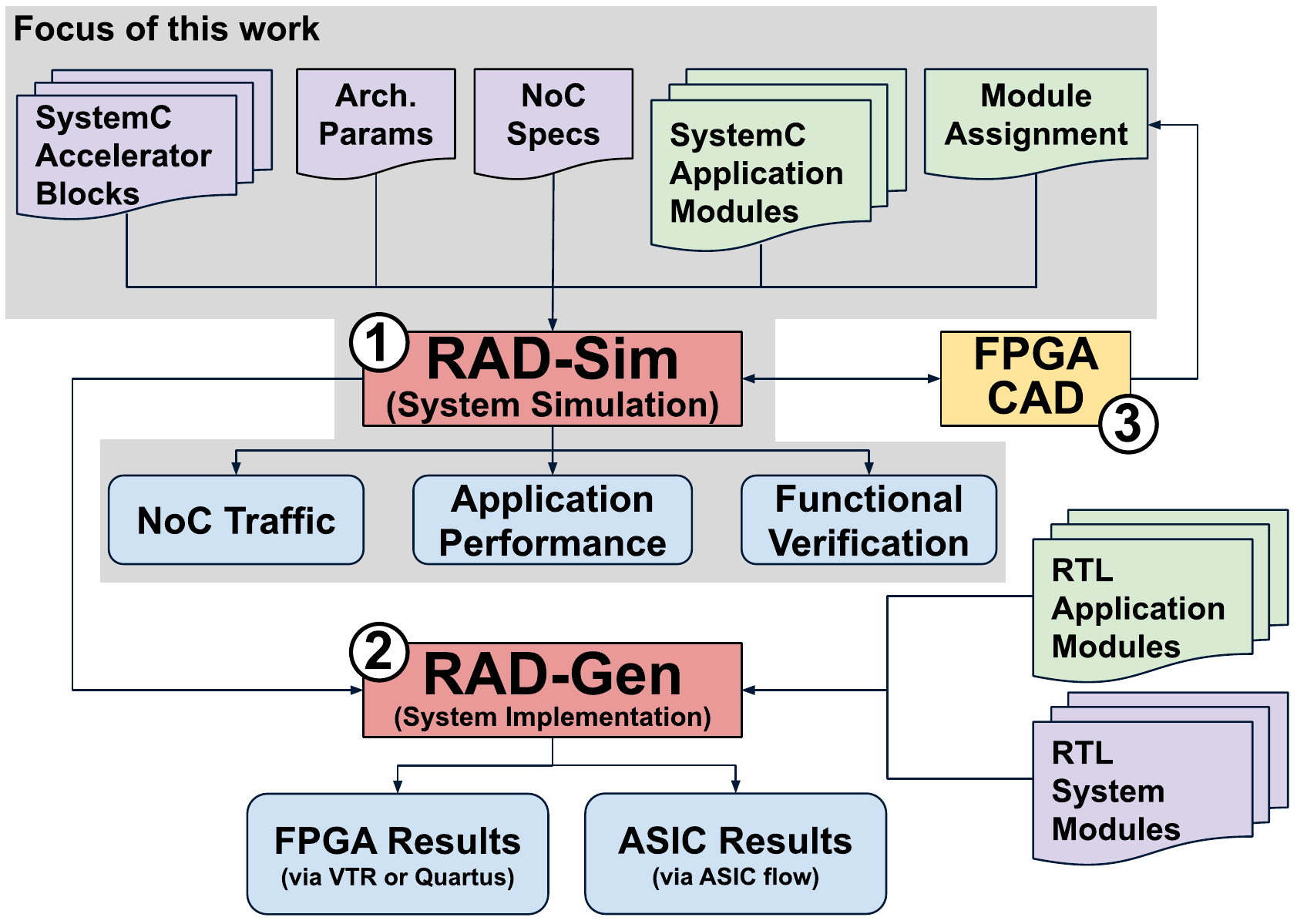}
    \caption{RAD architecture exploration and evaluation flow.}
    \label{fig:rad-flow}
    \vspace{-0.2cm}
\end{figure}

Fig. \ref{fig:rad-flow} shows an overview of our full RAD architecture evaluation flow, which consists of three main components.
The first component and the main focus of this paper is RAD-Sim, which allows rapid RAD design space exploration and evaluation of the interactions between design choices for different RAD components.
It takes as input a RAD architecture description in the form of architectural parameters, NoC specifications, and a set of SystemC models of the RAD's hard accelerator blocks.
In addition, it takes another set of SystemC models of application modules to be implemented on the FPGA fabric along with their assignment to specific NoC routers if they require access to the system-level NoC.
Then, it performs cycle-level simulation of the whole system to produce application performance results and NoC traffic reports.
It can also be used to verify the functionality of the application mapped to the specified RAD when provided with sets of test inputs and expected outputs.
This can be extremely useful when RADs and applications are co-designed during early stages of architecture exploration.

After RAD-Sim is used to rapidly narrow down the design space for target applications, more detailed evaluation can be performed for a few candidate RAD architectures using the second component of our flow, RAD-Gen.
This tool generates skeleton RTL code for the complete system including NoC routers, adapters, and module wrappers, in which the designer can drop in the RTL implementations of application modules and hard accelerator blocks.
Then, it pushes the portion of the design implemented on the programmable fabric through an FPGA CAD flow\footnote{VTR can directly model the embedded routers; to model them in Quartus we create reserved logic lock regions of the appropriate size and locations.} to get the design's maximum operating frequency and resource utilization.
It also pushes the NoC routers and any hard accelerator blocks through the ASIC implementation flow to get silicon area and timing results.

The third and final component of our flow is the link between conventional FPGA CAD tools and RAD-Sim. 
Hard NoCs on FPGAs present a new challenge for placement; modules must be placed not only where they have sufficient fabric resources and minimize traditional programmable routing, but also so that their connection to NoC adapters on nearby routers does not cause undue NoC congestion. 
RAD-Sim can act as an oracle for evaluating the connection of fabric modules to specific routers during placement. 
For example, the FPGA CAD tools can suggest a specific module assignment and pass it to RAD-Sim along with user-specified expected NoC traffic patterns.
RAD-Sim can then rapidly simulate this scenario and produce a report of expected latency for different traffic streams which the placement engine can use to adjust the module assignment and iterate again if latency constraints are not met.
This is analogous to invoking static timing analysis during the placement stage in the conventional FPGA CAD flow.
This work focuses only on the first component of our flow, RAD-Sim. 
The second and third components are in development and will be covered in future works.

\vspace{-0.1cm}
\subsection{RAD-Sim Implementation Details}
\vspace{-0.05cm}

RAD-Sim is developed in SystemC, which allows designers to model their hard accelerator blocks and application modules at various levels of abstraction, trading off model faithfulness for designer productivity.
For example, a specific module can be described using SystemC in a high-level behavioral way for fast development time, or a more detailed (closer to RTL) way that can be input to high-level synthesis tools to generate hardware.
RAD-Sim uses BookSim 2.0 \cite{jiang2013detailed} to perform cycle-accurate NoC simulation.
BookSim is an open-source NoC simulator that has been leveraged by many system simulators, such as GPGPU-Sim.
It is heavily parameterized to allow modeling a wide variety of interconnect networks with different topologies, routing functions, arbitration mechanisms, and router micro-architectures.

RAD-Sim builds on top of BookSim in three main aspects.
Firstly, RAD-Sim adds a SystemC wrapper around BookSim to allow designers to easily combine the NoC with different accelerator blocks and application modules modeled in SystemC.
Secondly, it complements BookSim by tracking packet contents to enable functional verification of actual applications on RADs. This is necessary because BookSim primarily focuses on performance estimation and hence models the arrival times of packets, not their contents. 
Finally, RAD-Sim also implements SystemC NoC adapters that allow RAD architects to experiment with different user-facing NoC abstractions, independently of the underlying NoC protocol.
These adapters also perform clock domain crossing and width adaptation between the application modules or hard accelerator blocks and the NoC.
For example, we provide users with AXI streaming (AXI-S) and AXI memory-mapped (AXI-MM) adapters, but RAD-Sim is structured to be modular such that architects can implement their custom or standardized NoC adapter protocol and easily integrate it in the simulator.

\begin{figure}[t!]
    \centering
    \includegraphics[width=0.9\linewidth]{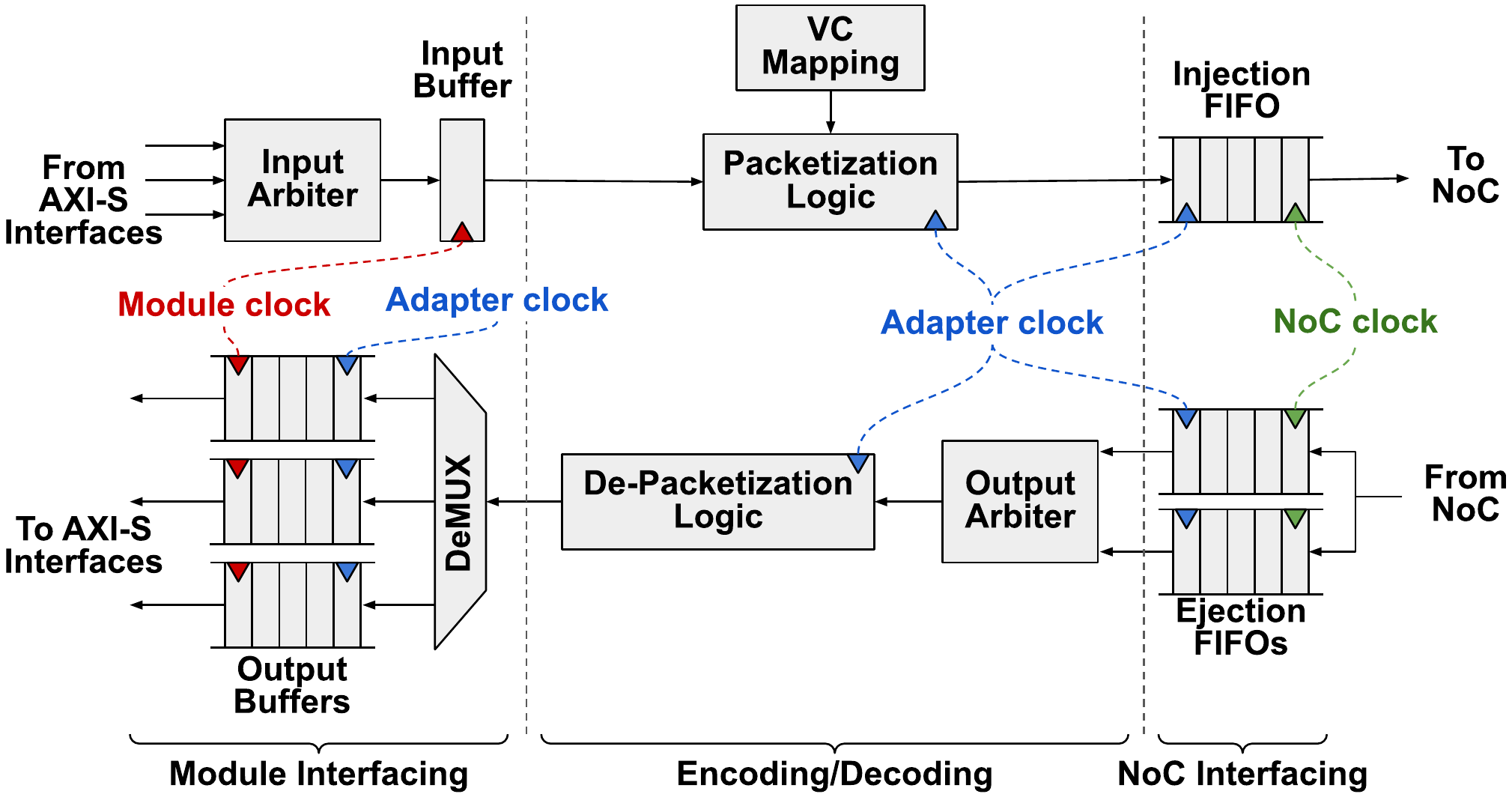}
    \caption{AXI-S slave (top) \& master (bottom) NoC adapters.}
    \label{fig:noc-adapters}
\end{figure}

Fig. \ref{fig:noc-adapters} shows the AXI-S master and slave NoC adapters implemented in RAD-Sim as an example.
They consist of three main stages: module interfacing, encoding/decoding, and NoC interfacing.
For the slave adapter, an input arbiter selects one of the (possibly multiple) AXI-S interfaces connected to the same NoC router.
Once a transaction is buffered, it is packetized into a number of NoC flits and mapped to a specific NoC virtual channel (VC).
Then, these flits are pushed into an asynchronous FIFO to be injected into the NoC depending on the router channel arbitration and switch allocation mechanisms.
The master adapter works in a similar way but in reverse: flits are ejected from the NoC and once a tail flit is received, they are depacketized into an AXI-S transaction which is then steered to its intended module interface. 
The adapters implemented in RAD-Sim are parameterized to allow experimentation with different arbitration mechanisms, VC mapping tables, and FIFO/buffer sizes. 
They also support up to three distinct clock domains where the connected module, adapter, and NoC are all operating at different clock frequencies.

\begin{table}[t!]
    \centering
    \caption{RAD-Sim architecture parameters.}
    \begin{tabular}{L{3cm} L{5.1cm}} 
    \hline\\ [-1.8ex]
    \textbf{User Input} & \textbf{Description} \\ [0.25ex] 
    \hline\\ [-1.8ex]
    \texttt{num\_nocs} &  No. of system-wide NoCs\\
    \texttt{noc\_payload\_width} & Bit width of NoC links for flit payload\\
    \texttt{noc\_freq} & NoC operating frequency\\
    \texttt{noc\_topology} & NoC topology (e.g. mesh, torus)\\
    \texttt{noc\_dim} & NoC dimensions (for certain topologies)\\
    \texttt{noc\_routing\_func} & NoC routing algorithm (e.g. XY, min hops)\\
    \texttt{noc\_vcs} & No. of NoC virtual channels\\
    \texttt{noc\_vc\_buffer\_size} & Depth of virtual channel buffers (words)\\
    \texttt{adapter\_interfaces} & No. of interfaces connected to each adapter \\
    \texttt{adapter\_fifo\_size} & Depth of adapter ejection/injection FIFOs\\
    \texttt{adapter\_obuff\_size} & Depth of adapter output buffer (words) \\
    \texttt{adapter\_in\_arbiter} & Adapter input arbitration mechanism\\
    \texttt{adapter\_out\_arbiter} & Adapter output arbitration mechanism\\
    \texttt{adapter\_vc\_mapping} & Mapping of flit types to virtual channels\\
    \texttt{adapter\_freq} & Adapter operating frequency\\
    \texttt{module\_freq} & Operating frequency for each module\\ 
    \texttt{num\_traces} & No. of event traces recorded\\
    \texttt{trace\_names} & Identifiers of recorded event traces \\ [0.25ex]
    \hline
    \end{tabular}
    \label{tab:knobs}
    \vspace{-0.2cm}
\end{table}

Table \ref{tab:knobs} lists some of the parameters that a user can tune to experiment with different RAD architectures.
Other more detailed NoC-specific options such as delay parameters, router micro-architecture, and switch/VC allocation mechanisms can also be specified directly using a BookSim configuration file.
In addition, RAD-Sim accepts as an input a module assignment file that specifies the NoC placement of all hard accelerator blocks and fabric modules (i.e. which NoC router each block/module port is connected to).
This is currently passed as a user-specified manual assignment. 
However, it can be automated to meet traffic latency constraints specified by the user or optimize the overall application performance.
As described in Sec. \ref{sec:flow-overview}, the FPGA CAD flow can potentially adjust the NoC placement of modules implemented on the FPGA fabric and invoke RAD-Sim to quantify the effect of these adjustments on the overall performance.

\begin{figure}[t!]
    \centering
    \includegraphics[width=0.9\linewidth]{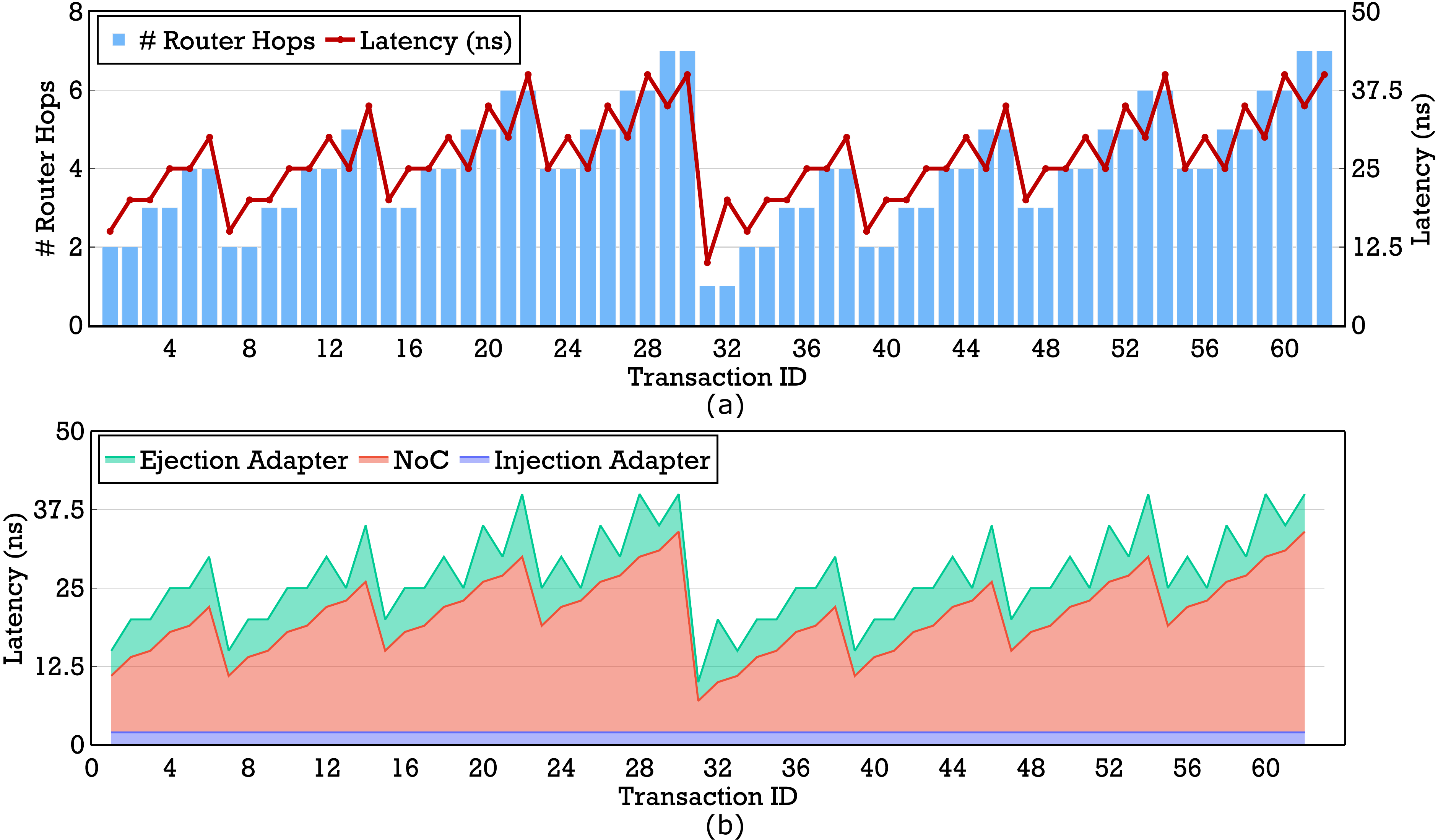}
    \caption{Example visualizations produced by RAD-Sim for an unloaded 4$\times$4 mesh NoC showing: (a) Overall communication latency, number of hops, and (b) Latency breakdown.}
    \label{fig:telemetry}
    \vspace{-0.2cm}
\end{figure}

In addition, RAD-Sim also provides telemetry utilities to record specific simulation events and traces along with different scripts to visualize the collected data.
This can be very useful in reasoning about the complex interactions between the different components of a RAD and understanding the effect of changing various architecture parameters on the overall system performance.
Fig. \ref{fig:telemetry} shows example visualizations produced by RAD-Sim when trying to characterize the unloaded communication latency for a RAD with a 4$\times$4 mesh NoC and two modules connected to each router.
In this example experiment, a single module sends two AXI-MM transactions to the first module connected to each router (15 routers $\times$ 2 transactions) one at a time, with no other traffic on the NoC. 
This then repeats for the second module connected to each router. 
The module, adapter and NoC operating frequencies are set to 200 MHz, 800 MHz, and 1 GHz, respectively.
The RAD-Sim telemetry utilities are used to record various timestamps in the transaction lifetime such as transaction initiation at the source module, packetization, injection/ejection, depacketization, and receipt at the destination module.
Fig. \ref{fig:telemetry}a shows the latency in nanoseconds and number of NoC router hops for each of the 62 issued transactions.
The graph shows how the number of hops and communication latency increase as the distance between the source and destination modules increases then drops when moving to the next row in the 4$\times$4 mesh of routers.
Fig. \ref{fig:telemetry}b shows another visualization produced by RAD-Sim that breaks down the latency for each transaction into time spent in the injection adapter, the NoC, and the ejection adapter.
This can highlight the overhead introduced when experimenting with different adapter implementations and protocols.

%% file: 04_npu.tex
\section{NPU Example Design}
\label{sec:npu}

\subsection{The Neural Processing Unit (NPU) Overlay}
\label{sec:npu-baseline}

For our study, we use the NPU overlay as a key benchmark from the DL application domain.
The NPU is a state-of-the-art FPGA soft processor for low-latency inference targeting memory-intensive DL models such as multi-layer perceptrons (MLPs), recurrent neural networks (RNNs), gated recurrent units (GRUs), and long short-term memory models (LSTMs).
It achieves state-of-the-art performance on Intel Stratix 10 NX FPGAs with DL-optimized tensor blocks.
On average, it achieves $24\times$ and $12\times$ higher performance than the same-generation Nvidia T4 and V100 GPUs, respectively~\cite{boutros2020beyond}.

\begin{figure}[t!]
    \centering
    \includegraphics[width=\linewidth]{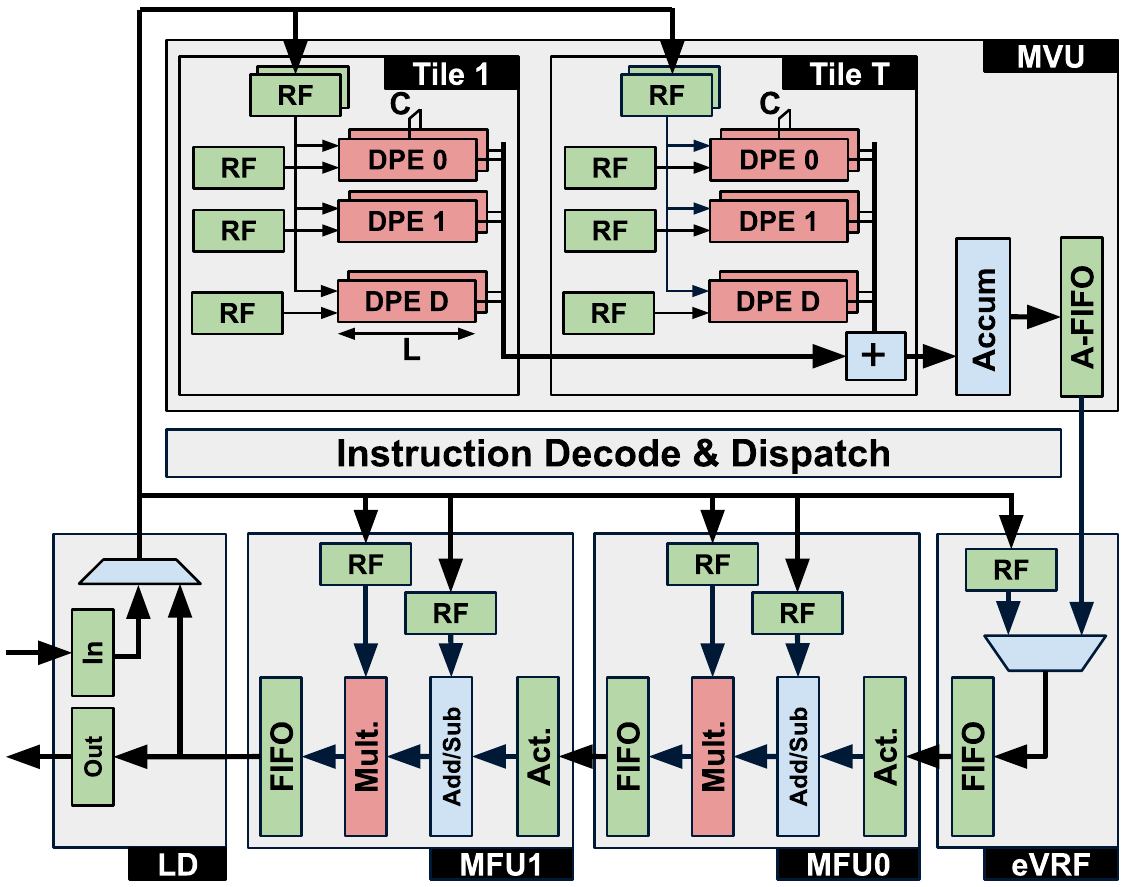}
    \caption{Overview of the NPU overlay architecture. The connections highlighted in red are latency sensitive channels.}
    \label{fig:npu-arch}
    \vspace{-0.2cm}
\end{figure}

Fig. \ref{fig:npu-arch} shows an overview of the NPU overlay architecture which consists of five chained blocks such that the outputs of one block are directly forwarded to the next.
The matrix-vector multiplication unit (MVU) consists of $T$ tiles, each of which has $D$ sets of $C$ dot-product engines (DPEs) of length $L$ multiplication lanes.
Each tile computes a portion of a matrix-vector multiplication operation, and then their partial results are reduced and accumulated over multiple time steps to produce the final MVU output.
This is followed by an external vector register file (eVRF) to skip the MVU for instructions that do not include a matrix-vector multiplication, and then two identical multi-function units (MFUs) for vector elementwise operations such as activation functions, addition/subtraction, and multiplication.
Finally, there is the loader block (LD) which writes back the pipeline results to any of the NPU's register files (RFs) and communicates with other system components (e.g. other modules or external interfaces).
All these blocks are orchestrated by very long instruction words that are decoded and dispatched to different blocks by a central control unit, as detailed in  \cite{nurvitadhi2019compete,boutros2020beyond}.

\subsection{Baseline SystemC NPU Model}
\label{sec:npu-systemc}

\begin{figure}[t!]
    \centering
    \includegraphics[width=\linewidth]{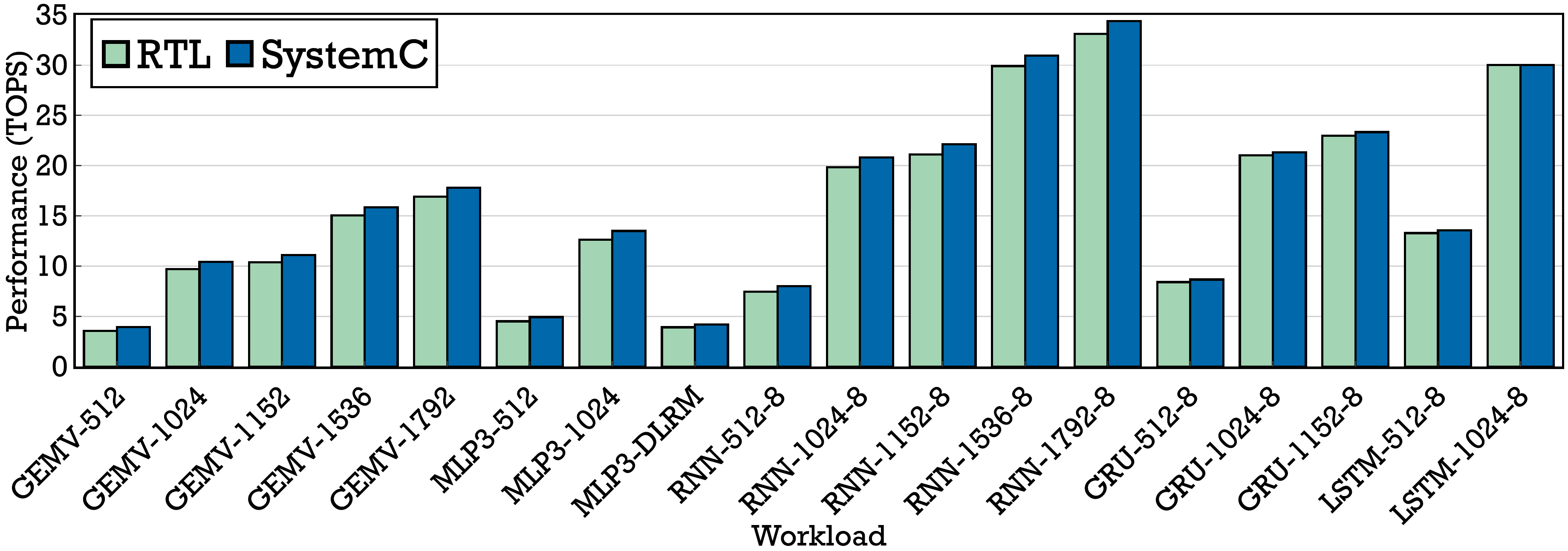}
    \caption{NPU performance results from RTL and SystemC simulations.}
    \label{fig:sim-compare}
    \vspace{-0.2cm}
\end{figure}

In order to use the NPU as a case study for RAD-Sim, we develop SystemC simulation models for its blocks such that we can later use them in RAD-Sim as either hard accelerator blocks or fabric application modules.
These models are parameterized such that we can experiment with different NPU architecture parameters ($T, D, C$ and $L$) and module latencies depending on their low-level implementation details.
To evaluate the speed and accuracy of our NPU SystemC simulation model, we compare it to cycle-accurate RTL simulation of the NPU SystemVerilog implementation.
For our experiments, the RTL simulation uses Synopsys VCS v2016.06, and both the SystemC and RTL simulations are performed on the same Intel Xeon Gold 6146 24-core CPU.
We use an NPU configuration similar to that in \cite{boutros2020beyond} with 2 cores, 7 tiles, 40 DPEs and 40 lanes, which we also use for the rest of our experiments in this paper.
We run simulations for a variety of NPU workloads including simple matrix-vector multiplications (GEMV), RNNs, GRUs, LSTMs, and MLPs of different sizes, and report the results in Fig. \ref{fig:sim-compare} in tera operations per second (TOPS).
The results show that our SystemC simulation model can estimate NPU performance to a high degree of accuracy with average error of only 5.1\% and maximum error of 10.8\% compared to cycle-accurate RTL simulation.
However, the SystemC simulations are 26$\times$ faster than the RTL simulations on average, with speedups ranging from 6.5$\times$ to 100$\times$ depending on the workload size.
This highlights the significant speed difference between SystemC and  RTL simulation which is a key pillar of RAD-Sim and builds confidence in the performance estimates that we generate using this NPU model for the rest of our experiments.

\subsection{Mapping and Simulating the NPU on a RAD Instance}
\label{sec:npu-rad}

We modified the NPU to use latency-insensitive interfaces so we are able to connect them via the system-level NoC of a RAD instance.
This completely decouples the application compute from its inter-module communication, and raises the interconnect abstraction level enabling the exploration of complex RADs that incorporate hard accelerator blocks.
In this case, the conventional FPGA CAD tools do not need to optimize the timing and routability of signals crossing module boundaries or trying to reach the programmable routing interfaces of a hard accelerator block.
If each application module meets timing separately and can be connected to a NoC adapter, the evaluation of end-to-end application performance on a given RAD instance is raised to the cycle-level simulation of soft/hard modules and NoC latency; this is exactly what is captured by RAD-Sim.

We map the NPU to an example RAD instance with an FPGA fabric and a separate complex of hard accelerator blocks, as shown in Fig.~\ref{fig:rad-instance}, and evaluate its overall performance using RAD-Sim.
In this case, we implement matrix-vector multiplication units that resemble the MVU tiles of the NPU (see Fig.~\ref{fig:npu-arch}) as the hard accelerator blocks that can only be accessed from the fabric via the NoC.
These blocks are realistic candidates for hardening since they implement common functionality across almost all DL workloads, while the rest of the NPU blocks could be specialized for different workloads to increase efficiency \cite{boutros2021specializing} and thus benefit from the FPGA's reconfigurability.

We define the term \textit{FPGA sector} as a region of FPGA resources with a NoC router/adapter at its center.
For example, an FPGA with 8$\times$5 sectors has a total of 40 NoC routers/adapters throughout its fabric.
Equivalently, we define an \textit{ASIC sector} as an area of silicon that has the same footprint of an FPGA sector and includes a hard accelerator block (possibly with other hardened components) and a NoC router.
The example RAD instance that we use in this experiment has an 8$\times$5 grid of FPGA sectors and a 2$\times$5 side complex of ASIC sectors.
The FPGA sectors collectively have the same resources as our baseline Intel Stratix 10 NX 2100 device ($702$k ALMs, $6,847$ BRAMs, $3,960$ tensor blocks).


We map the NPU to our example RAD instance and evaluate its performance using RAD-Sim.
We set an FPGA fabric operating frequency of 300 MHz (matching the NPU operating frequency in \cite{boutros2020beyond}) and conservatively assume that the hard accelerator blocks run only at 600 MHz.
We scale the operating frequency of the 28nm NoC routers from \cite{abdelfattah2015take} to 1.5 GHz in the Stratix 10 14nm process technology, and we assume that the NoC adapters operate at 4$\times$ the fabric speed, similarly to \cite{abdelfattah2015take}.
In our experiments, we use a mesh NoC topology with dimensions equal to the total number of FPGA and ASIC sectors (i.e. 10$\times$5 mesh) with 3 VCs and dimension order routing.
The depths of the NoC adapter injection/ejection FIFOs and ouptut buffers (see Fig. \ref{fig:noc-adapters}) are set to 16 and 2, respectively.
We manually assign the NPU vector elementwise modules (eVRF, MFUs, LD, Insruction Dispatcher) implemented on the FPGA fabric to specific NoC routers in a reasonable (but possibly sub-optimal) placement.

\subsection{Implementation Results}

\begin{table}[t!]
    \centering
    \caption{Resource utilization for the NPU modules implemented on the RAD FPGA fabric.}
    \begin{tabular}{C{2.2cm} C{2.2cm} C{2.2cm}} 
    \hline\\ [-1.8ex]
    \textbf{ALMs} & \textbf{BRAMs} & \textbf{Tensor Blocks}\\ [0.25ex] 
    \hline\\ [-1.8ex]
    550,0930 (78\%) & 2,632 (90\%) & 3,200 (81\%)\\ [0.25ex]
    \hline
    \end{tabular}
    \label{tab:fpga-resources}
\end{table}

To determine FPGA resource utilization, we synthesize, place and route the NPU modules mapped to the FPGA fabric using Intel Quartus Prime Pro 21.2 on a Stratix 10 NX 2100 device.
We use reserved logic lock regions at the appropriate locations for NoC routers and adapters, mark them as empty design partitions, and connect the NPU modules to them based on our manual module assignment to different routers. 
We conservatively size each logic lock region as a grid of 10$\times$10 logic array blocks (LABs) compared to the 3$\times$3 LAB region used in \cite{abdelfattah2012design}, as we are using 128-bit wide links vs. the 32-bit wide links of \cite{abdelfattah2012design}.
Table \ref{tab:fpga-resources} shows the resource utilization of the NPU modules implemented on the FPGA fabric.

We also verify that the matrix-vector multiplication units we chose to implement as hard accelerator blocks fit in the available ASIC sector area footprint using FPGA resources silicon areas and FPGA-to-ASIC area scaling ratios from \cite{wong2011comparing}, \cite{boutros2018you} and \cite{boutros2018embracing}.
Our estimates show that the hard matrix-vector unit consumes less than 55\% of the available ASIC sector area leaving more than enough area for the NoC routers, adapters, links, and any additional hardened functionality.
In the future, the RAD-Gen component of our flow, described in Sec. \ref{sec:flow-overview}, will automate any manual steps needed to obtain the FPGA results and will push the RTL implementation of the hard accelerator blocks through the ASIC design flow to obtain exact area and timing results.

\subsection{Performance Results}

\begin{figure}[t!]
    \centering
    \includegraphics[width=\linewidth]{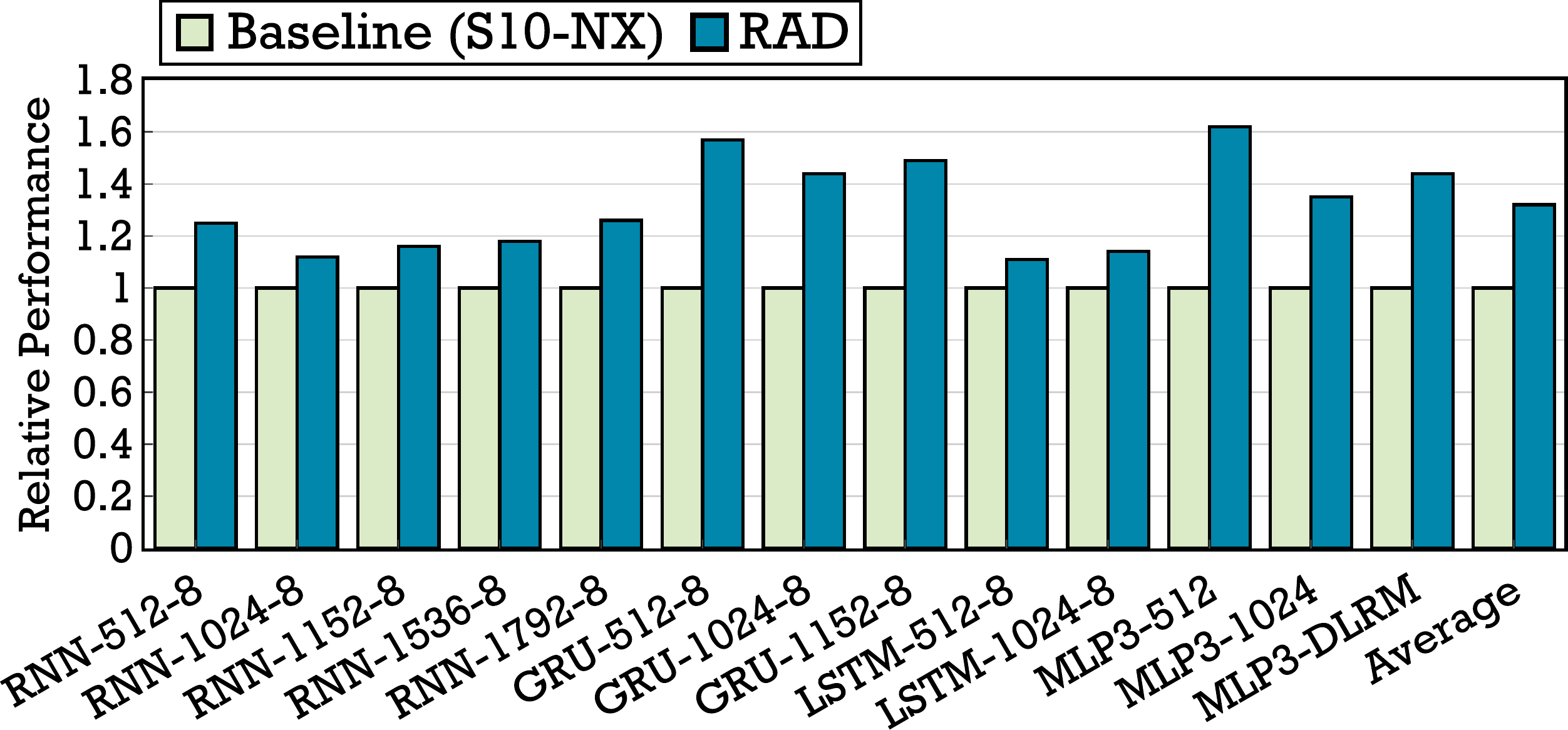}
    \caption{Relative performance comparison of the NPU on Stratix 10 NX and our example RAD instance.}
    \label{fig:sim-compare}
    \vspace{-0.2cm}
\end{figure}

\begin{figure}[t!]
    \centering
    \includegraphics[width=\linewidth]{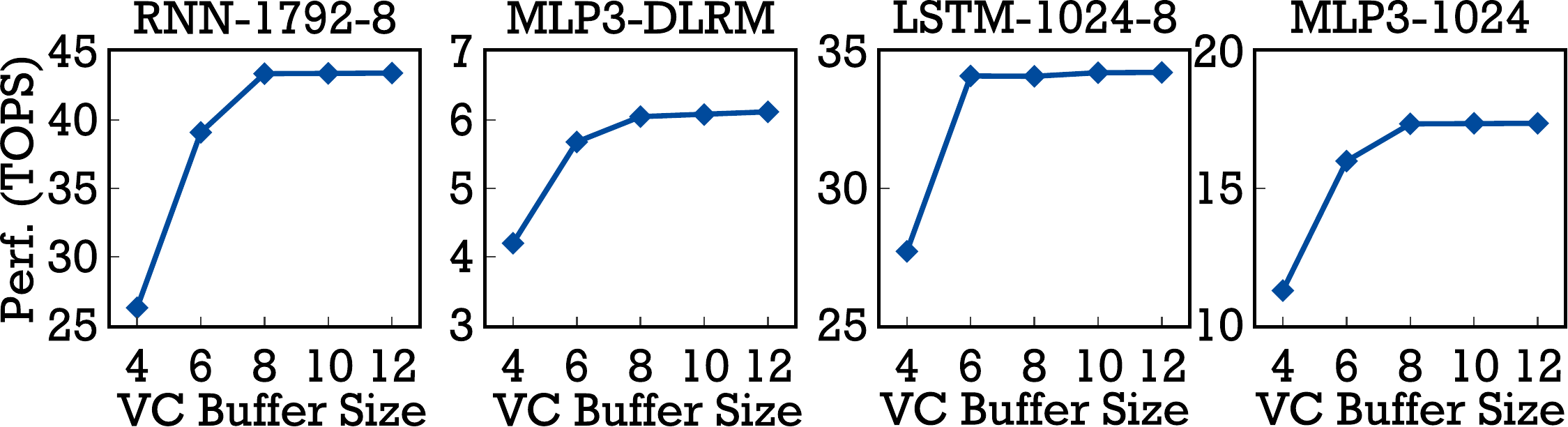}
    \caption{Effect of changing NoC VC buffer size on NPU performance for select workloads.}
    \label{fig:vc-results}
    \vspace{-0.2cm}
\end{figure}

Fig. \ref{fig:sim-compare} shows the relative performance comparison between the baseline NPU on Stratix 10 NX from \cite{boutros2020beyond} and that when mapped to our example RAD instance.
The NPU implemented on the RAD achieves, on average, 1.32$\times$ higher performance compared to the baseline conventional Stratix 10 NX by exploiting the hardened MVU coarse-grained accelerator blocks and instantiating more vector elementwise engines in soft logic using the freed up FPGA fabric resources.
RAD-Sim also enables us to study the effect of different choices of architecture parameters on the end-to-end application performance.
Fig. \ref{fig:vc-results} shows the impact of changing the VC buffer size in the NoC routers of our example RAD instance. 
VC buffers with depth less than 8 flits can throttle performance given the NPU traffic patterns when using the specified NoC specifications and placement of NPU modules.
On the other hand, VC buffer depths of more 8 flits yield minimal or no additional performance benefits.